# Detuning the Honeycomb of the α-RuCl$_3$ Kitaev lattice: A case of Cr$^{3+}$ dopant

*Maria Roslova*[\*,†,1]*, Jens Hunger*[1]*, Gaël Bastien*[2]*, Darius Pohl* [2,3]*, Hossein M. Haghighi*[2]*, Anja U. B. Wolter*[2]*, Anna Isaeva*[1]*, Ulrich Schwarz*[4]*, Bernd Rellinghaus*[3]*, Kornelius Nielsch*[2]*, Bernd Büchner*[2]*, and Thomas Doert*[1]

1 Faculty of Chemistry and Food Chemistry, Technische Universität Dresden, 01062 Dresden, Germany

2 Institute for Solid State and Materials Research (IFW) Dresden, 01171 Dresden, Germany

3 Dresden Center for Nanoanalysis, cfaed, TU Dresden, 01062 Dresden, Germany

4 Max Planck Institute for Chemical Physics of Solids, 01187 Dresden, Germany

Abstract

Fine-tuning chemistry by doping with transition metals enables new perspectives for exploring Kitaev physics on a two-dimensional (2D) honeycomb lattice of *α*-RuCl$_3$, which is promising in the field of quantum information protection and quantum computation. The key parameters to vary by doping are both Heisenberg and Kitaev components of the nearest-neighbor exchange interaction between the $J_{\text{eff}} = ½$ Ru$^{3+}$ spins, depending strongly on the peculiarities of the crystal structure. Here, we successfully grew single crystals of the solid solution series Ru$_{1-x}$Cr$_x$Cl$_3$ ($0 \leq x \leq 1$) with Cr$^{3+}$ ions coupled to the Ru$^{3+}$ Kitaev host using chemical vapour transport reaction. The Cr$^{3+}$ substitution preserves the honeycomb type lattice of α-RuCl$_3$ with mixed occupancy of Ru/Cr sites, no hints on cationic order within the layers were found by single crystal X-ray diffraction and transmission electron microscopy investigations. In contrast to the high quality single crystals of *α*-RuCl$_3$ with ABAB ordered layers, the ternary compounds demonstrate a significant stacking disorder along the *c*-axis direction evidenced by X-ray diffraction and high resolution scanning transmission electron microscopy (HR-STEM). Raman spectra of substituted samples are in line with a symmetry conservation of the parent lattice upon chromium doping. At the same time, magnetic susceptibility data indicate that the Kitaev physics of *α*-RuCl$_3$ is increasingly repressed by the dominant spin-only driven magnetism of Cr$^{3+}$ in Ru$_{1-x}$Cr$_x$Cl$_3$.

**Introduction**

A critical aspect for advancing knowledge and practical applications of complex materials is the ability to control their electronic properties via chemistry routes. Chemical doping, when the substituting atoms become an integral part of the electron system, is a powerful tool to trigger and vary electronic interactions at the atomic level, which makes it especially advantageous for 2D systems, for example, 5$d$ and 4$d$ oxides and halides hosting new types of magnetic ground states and excitations, such as Kitaev spin liquids[1]. The latter stimulated great interest due to their potential to protect quantum information[2,3] or to provoke the emergence of Majorana fermions.[4,5,6] Due to peculiarities of the crystal structure, α-RuCl$_3$ is currently under consideration as a candidate for a Kitaev−Heisenberg system describing the competition of bond dependent magnetic exchange interactions in a honeycomb lattice structures.[1] α-RuCl$_3$ consists of very weakly bonded layers of edge-sharing RuCl$_6$ octahedra with the central Ru$^{3+}$ (4$d^5$) ions forming an almost ideal honeycomb arrangement, which is essential for the Kitaev−Heisenberg model. The stacking sequence of the layers may differ depending on synthesis conditions which lead to the symmetry lowering observed in X-rays and neutron diffraction measurements.[7,8,9,10,11] Single crystals of α-RuCl$_3$ with minimal stacking faults are seen to consistently exhibit a monoclinic unit cell (space group $C2/m$)[12] corresponding to a stacking of layers similar to those in AlCl$_3$ [13] or iridates[14,15] (ABAB stacking). So far, little is known experimentally about the nature of the phases that derive from chemical doping of α-RuCl$_3$. Very recently the series Ru$_{1−x}$Ir$_x$Cl$_3$ has been investigated on crystals and polycrystalline samples. [16] Low-spin 5$d^6$ Ir$^{3+}$ may be considered as a non-magnetic impurity in the $J_{eff}$ = 1/2 Ru$^{3+}$ magnetic sublattice while the identical ionic radii of Ru$^{3+}$ and Ir$^{3+}$ should preserve a regular $M$Cl$_6$ environment. Interestingly, iridium doping does not lead to complete magnetic order suppressing although the Neel temperature is found to be shifted towards lower temperatures with increasing dopant content. Neutron diffraction experiment reveal Ru$_{1−x}$Ir$_x$Cl$_3$ crystals to exhibit the same magnetic Bragg peaks as disordered large crystals of α-RuCl$_3$, corresponding to a mixture of ABAB and ABCABC stacking periodicity. [16] Moreover, the volume fractions of these phases vary from one sample to the next. Other mixed-metal trihalides Ru$_{1−x}M_x$Cl$_3$ have received little attention in this respect so far.

The crystal structure of the layered monoclinic transition metal trihalides can be regarded as a distorted variant of the anionic hexagonal close-packing with 2/3 of the octahedral voids filled by the transition metal cations in every second layer. The deviation from the ideal structure is typically described by two criteria – the dislocation of the metal atom layers from the perfect hexagonal net and the perturbation of the dense anionic close-packing. The former can be caused by emergence of intra-layer metal-metal interactions, whereas the latter is related to polarization of anions by the highly-charged transition metal cations. According to these criteria, the crystal structures of $CrCl_3$, $AlCl_3$, $IrCl_3$ and $RhCl_3$ are most closely related to α-$RuCl_3$.

The $Ru_{1-x}M_xCl_3$ ($M$ = Cr, Ir, Rh) compounds were prepared for the first time more than 20 years ago by deposition from the gas phase.[17,18] The series of chromium doped crystals were used as a model system for the direct determination of the dopant content in the $RuCl_3$ structure by scanning tunneling microscopy (STM).[19] Crystals of compositions $Ru_{1-x}Cr_xCl_3$ with $x$ = 0.11, 0.15 and 0.20 were reported.[19] The positions of the chromium atoms are found to be statistically distributed in the (001) face by means of STM for the $Ru_{0.8}Cr_{0.2}Cl_3$ sample. The single crystal X-ray diffraction (XRD) based characterization was not provided due to low quality of the crystals containing twinning and a considerable amount of stacking faults. Up to date, the electronic and magnetic properties as well as bulk crystal structure of $Ru_{1-x}Cr_xCl_3$ are not investigated. Due to the close structural proximity to α-$RuCl_3$, the regular $MCl_6$ environment is preserved, but in contrast to α-$RuCl_3$, $CrCl_3$ represents a material with predominant Heisenberg interactions. $CrCl_3$ undergoes magnetic ordering below at $T_N$ = 17 K and recent reports indicate a two-step ordering process[20]. Neutron diffraction investigations at 4.2 K [21] revealed that the magnetic moments lie in the honeycomb (001) plane with adjacent ferromagnetic layers being aligned in an antiparallel way.

Thus, $3d^3$ $Cr^{3+}$ is expected to be a magnetic impurity in the $J_{eff} = 1/2$ $Ru^{3+}$ sublattice and may be regarded as a first experimental system representing $3d$ magnetic impurities coupled to a Kitaev Heisenberg magnet.

Here we report the synthesis of the full substitution series $Ru_{1-x}Cr_xCl_3$ ($0 \leq x \leq 1$) as well as crystal growth, structure determination and electron microscopy investigations at the atomic level of samples with $x = 0.05$, $0.1$ and $0.5$. Finally, the evolution of the magnetic properties in the series is characterized by magnetization measurements.

**Experimental**

All preparation steps were performed in an argon-filled glovebox with $O_2$ and $H_2O$ content less than 0.1 ppm. For the synthesis, pure ruthenium powder (-325 mesh, 99.9% Alfa Aesar) and chromium powder (-100 mesh, 99.5% Aldrich) were filled into a quartz ampoule, together with a sealed silica capillary containing chlorine gas (99.5 % Riedel-de Haën). The chlorine gas was dried prior to use by bubbling through conc. $H_2SO_4$ and a $CaCl_2$ column, the ruthenium and chromium powders were used was without further purification. The molar ratio of the starting materials was chosen with a slight excess of chlorine to ensure in-situ formation of $RuCl_3$ and $CrCl_3$ and their chemical transport according to the reactions:

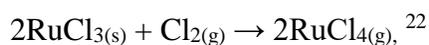

$2RuCl_{3(s)} + Cl_{2(g)} \rightarrow 2RuCl_{4(g),}$ [22]

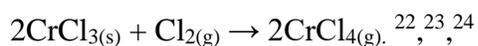

$2CrCl_{3(s)} + Cl_{2(g)} \rightarrow 2CrCl_{4(g).}$ [22],[23],[24]

The starting compositions for different samples are given in Table 1. After sealing of the reaction ampoule under vacuum, the chlorine-containing capillary inside was broken in order to release the gas. The ampoule was kept in a two-zone furnace with the temperature gradient between 750 °C (source) and 650 °C (sink) for 5 days. It was found that a prolongation of the growth time had little effect on both the number and size of the crystals. A successful growth exhibits few well separated platelike black crystals as displayed in Figure 1(a). These crystals have individual facets of up to 0.8 cm$^2$ in area. No oxides and no elemental metal phases were identified in the powder patterns of the obtained mixed phases. Insignificant amounts of green $Cr_2O_3$ powder were found at the source sides from time to time but this phase has never been transferred to the sink.

X-ray powder diffraction (PXRD) patterns were collected using a PANalytical X'Pert Pro MPD diffractometer with Cu$K\alpha_1$ radiation ($\lambda$ = 1.54056 Å) at room temperature in the 2$\theta$ range between 5 and 90 °, with a scan speed of 0.01 ° per second and a step size of 0.026 °.

Single crystals intensity data were collected at $T$ = 100(1) K using either, a four-circle diffractometer *Supernova* (Rigaku-Oxford Diffraction) equipped with a hybrid photon counting detector or an Apex II (Bruker-AXS) with a CCD detector, with graphite-monochromated Mo-$K\alpha$ radiation ($\lambda$ = 71.073 pm). Data were corrected for Lorentz and polarization factors, and a multi-scan absorption correction was applied. [25,26] The crystal structures were refinement against $F_o^2$ were performed with JANA2006 [27] using the site parameters of $\alpha$-RuCl$_3$ as start model.

The SEM images were collected on a Hitachi SU8020 microscope equipped with a field-emission gun. An acceleration voltage of 2 kV and a current of 10 µA were used to generate the SEM images in the secondary electron scanning mode. The energy dispersive X-ray spectra (EDX) were collected using an Oxford Silicon Drift X-Max$^N$ detector at an acceleration voltage of 10 or 20 kV and with 100 s accumulation time.

Selected area electron diffraction (SAED) patterns and bright-field (BF) TEM images were recorded with a FEI Tecnai-T20 transmission electron microscope (with a thermo-emission $LaB_6$ cathode, operated at 200 kV) equipped with EDX analyzer. For a specimen preparation the crystals of $Ru_{0.91(4)}Cr_{0.10(2)}Cl_3$ were ground in an agate mortar and then were dispersed in isopropanol following by ultrasonic bath treatment for five minutes. A few drops of the solution were put on a copper grid with a holey carbon film. The homogeneity of microcrystallites composition was confirmed by energy dispersive X-ray (EDX) analysis using the Ru-*L*, Cr-*K*, and Cl-*K* lines. The Tecnai imaging and analysis (TIA) software was used to record images with the digital camera.

High resolution Scanning transmission electron microscopy (HR-STEM) have been performed using FEI Titan$^3$ 80-300 probe and image corrected microscope operated at 300 kV. The convergence semi-angle was set to 21.5 mrad. Fast HR-STEM image series had been acquired and post processed using SmartAlign by HREM Research. [28] For the HR-STEM study, a parallel-sided lamellae was micromachined from the $Ru_{0.91(4)}Cr_{0.10(2)}Cl_3$ crystal perpendicular to its *ab*-plane using a 30 keV focused ion beam, FIB Crossbeam 1540 XB, Zeiss. At the final stage of specimen preparation, at a thickness of approximately 600 nm, low 120 keV cleaning was carried out using 2 keV FIB milling to reduce the effects of specimen surface damage and Ga implantation. The final lamellae composition was also confirmed by EDX (see Supporting Information, Figure S1).

Raman spectra were recorded with a LabRam System 010 (Jobin Yvon) in backscattering mode. The setup, equipped with a microscope (objectives 10× and 50×) and additional filters for low-frequency performance, used the He–Ne 633 nm line with 15 mW as excitation source. To prevent radiation- or heat-induced oxidation processes on the crystal surfaces, the laser power was attenuated and crystals were mounted in quartz capillaries under argon. The capillaries were then flame sealed and fixed on a glass slide.

The magnetization was measured using a superconducting quantum interference magnetometer SQUID-5T "Quantum Design" in a temperature range of $2 \leq T \leq 300$ K at $\mu H = 0.1$ and 1 T applied parallel to the *ab*-plane of a crystal after zero-field cooling.

**Results and Discussion**

$Ru_{1-x}Cr_xCl_3$ crystals exhibit a platelike morphology with flat, shiny, black surface as visible from Figure 1a; crystals with chromium content higher that $x \geq 0.8$ are dark brown, pure $CrCl_3$ is purple. The typical SEM micrographs of a single crystal are presented in Figure 1, b–d. No obvious morphology change in $Ru_{1-x}Cr_xCl_3$ was detected as compared to pure *α*-$RuCl_3$ [29], Figure 1a;[30], Figure 1b. The largest crystal faces correspond to the (001) face and in the orthogonal direction terrace-like structures with steps of a few micrometers thickness can be seen. The individual steps and layers are shifted and rotated relative to each other without obvious coherence (Figure 1d). The crystals growth most likely proceeds by a layer-by-layer deposition including the formation of flat, low-growth rate facets, as it was observed previously for other layered compounds with a high volatility. [31],[32],[33] Due to the highly anisotropic morphology the $Ru_{1-x}Cr_xCl_3$ crystals selectively cleave along the 100 direction. Mechanical manipulation of the $Ru_{1-x}Cr_xCl_3$ crystals can, like in the case of α-$RuCl_3$, easily introduce more layer misorientation and stacking faults, resulting in a higher amount of diffuse intensity in the diffraction images. The homogenous distribution of Ru and Cr in the samples is shown on a crystal of $Ru_{0.91(4)}Cr_{0.10(2)}Cl_3$ as example in Figure 2.

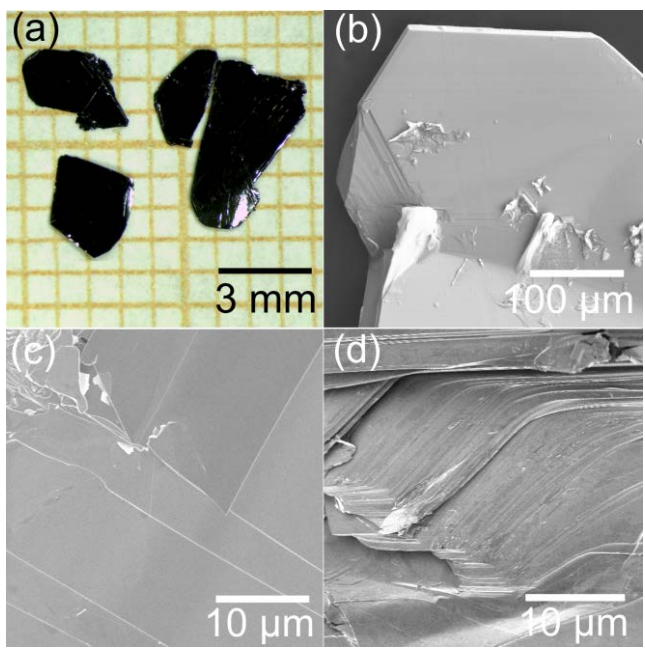

**Figure 1.** (a) $Ru_{1-x}Cr_xCl_3$ crystals grown by chemical vapor transport (b–d) Scanning electron micrographs taken at 2.0 keV in the SE mode showing the morphology of the $Ru_{1-x}Cr_xCl_3$ crystals in the *ab*-plane and along the stacking fault direction.

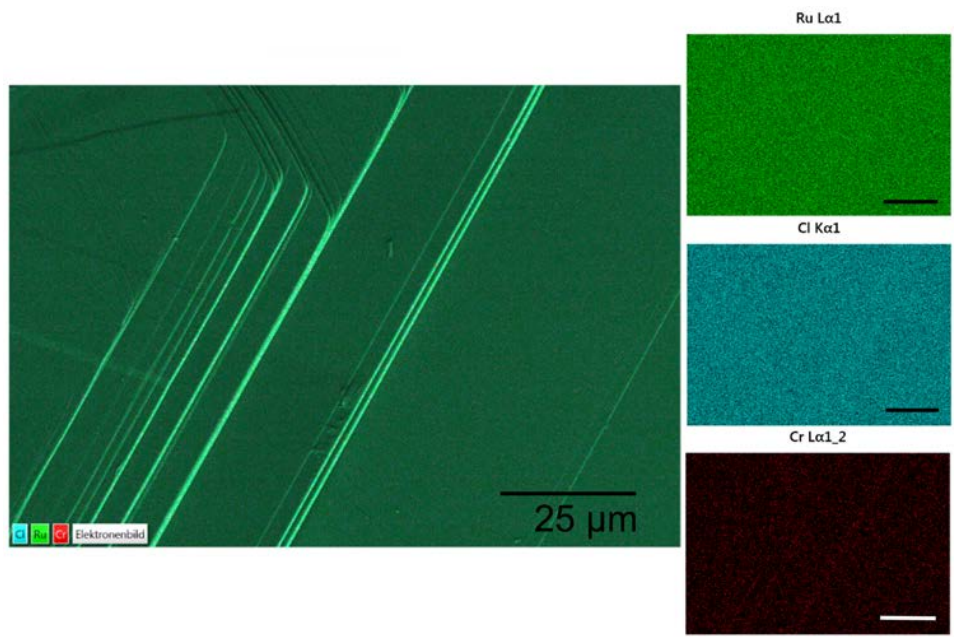

**Figure 2.** Secondary electron (SE) image and EDX color maps from a representative Ru/Cr mixed metal $Ru_{0.91(4)}Cr_{0.10(2)}Cl_3$. The scale bar corresponds to 25 µm.

Very recently the thermodynamics of α-RuCl$_3$ crystal growth was modelled by CalPhaD method [34] with a focus to production of the α-RuCl$_3$ nanosheets with a thickness of about 20 – 200 nm. With respect to the calculation results [34], the transport of α-RuCl$_3$ is possible without any additional transport agent. Our experiments confirm that more compositionally complex Ru$_{1-x}$Cr$_x$Cl$_3$ phases can be also deposited from a vapor phase without changes in their composition even in the absence of Cl$_2$ transport agent.

**X-ray diffraction**

X-ray powder diagrams indicated phase pure products for the complete series Ru$_{1-x}$Cr$_x$Cl$_3$; no peak splitting was observed which might point towards structural changes or phase separation. To investigate the trend of the lattice parameters in the Ru$_{1-x}$Cr$_x$Cl$_3$ series, we measured X-ray diffraction patterns for all compounds with a Si standard (see Supporting Information, Figure S2). The powder patterns indicate an extremely strong texture, (00L) reflections are dominant compared to other observed diffracted intensities. The patterns were indexed in the monoclinic $C2/m$ unit cell of the binary border phases (α-RuCl$_3$ for samples with $x \leq 0.5$ and CrCl$_3$ for samples with $x > 0.5$) and the lattice parameters were refined by Rietveld analysis. As can be seen from Figure 3d, the lattice parameter $c$ – i.e. the stacking direction of the honeycomb $M$Cl$_3$-layers of the mixed crystals – increases with increasing chromium content. In the *ab*-plane the substitution of Ru by Cr leads to a compression of the α-RuCl$_3$ honeycomb net along the *b*-axis, whereas the *a* lattice parameter remains almost constant. The monoclinic angle *β* decreases slightly, from 108.96(1) ° in Ru$_{1.00(2)}$Cr$_{0.019(1)}$Cl$_3$ to 108.76(1) ° in Ru$_{0.31(1)}$Cr$_{0.74(1)}$Cl$_3$ (for comparison, the monoclinic angles in RuCl$_3$ and CrCl$_3$ are 108.96(1) ° and 108.49(1) °, correspondingly). It is interesting to note that these tendencies differ from those found for the $Ru_{1-x}Ir_xCl_3$ substitution series [35]. Despite the smaller crystal ionic radii of $Cr^{3+}$ (75.5 pm; high-spin configuration) as compared to $Ru^{3+}$ and $Ir^{3+}$ (82 pm), all in octahedral coordination [36], the volume of the unit cell increases by both Cr and Ir doping in comparison to pure α-$RuCl_3$. Thus, for the maximal achievable Ir content ($x = 0.35$ [16]) the volume of the unit cell is ~0.4 % larger than that of α-$RuCl_3$ while for the comparable substitution degree of Cr ($x = 0.37$) the increase of the unit cell volume is about 0.5 %. In contrast to $Ru_{1-x}Ir_xCl_3$, the $Ru_{1-x}Cr_xCl_3$ solid solution samples do not strictly obey Vegard's law even in the low Cr-substitution regime. The starting (nominal) compositions of $Ru_{1-x}Cr_xCl_3$ samples and the composition obtained by means of the EDX analysis on single crystals are summarized in Table 1 and Figure 3 together with the results of the PXRD analysis. It is also worth noticing that X-ray peak intensities, EDX results as well as the physical appearance of crystals remained unchanged after storing the crystals several weeks under ambient conditions, demonstrating stability of the chromium doped samples in air.

**Table 1.** Starting compositions, EDX results and PXRD-derived lattice parameters for $Ru_{1-x}Cr_xCl_3$ crystals

| Starting composition | EDX | $a$, Å | $b$, Å | $c$, Å | $\beta$, ° | $V$, Å$^3$ |
|---|---|---|---|---|---|---|
| $RuCl_3$ | $RuCl_3$ | 5.9719(3) | 10.3636(5) | 6.0466(2) | 108.960(4) | 353.92(1) |
| $Ru_{0.98}Cr_{0.02}Cl_3$ | $Ru_{1.00(2)}Cr_{0.019(1)}Cl_3$ | 5.9731(3) | 10.3639(4) | 6.0428(4) | 108.967(5) | 353.76(2) |
| $Ru_{0.95}Cr_{0.05}Cl_3$ | $Ru_{0.96(1)}Cr_{0.05(1)}Cl_3$ | 5.9573(6) | 10.3676(6) | 6.0571(5) | 108.989(9) | 353.15(4) |
| $Ru_{0.9}Cr_{0.1}Cl_3$ | $Ru_{0.91(4)}Cr_{0.10(2)}Cl_3$ | 5.9597(4) | 10.3567(5) | 6.0456(3) | 108.695(6) | 353.46(2) |
| $Ru_{0.8}Cr_{0.2}Cl_3$ | $Ru_{0.84(2)}Cr_{0.16(1)}Cl_3$ | 5.9660(4) | 10.365(1) | 6.0644(6) | 109.05(1) | 355.15(3) |
| $Ru_{0.75}Cr_{0.25}Cl_3$ | $Ru_{0.76(1)}Cr_{0.24(1)}Cl_3$ | 5.9721(3) | 10.3369(4) | 6.0714(3) | 109.00(5) | 354.40(1) |
| $Ru_{0.7}Cr_{0.3}Cl_3$ | $Ru_{0.69(1)}Cr_{0.37(1)}Cl_3$ | 5.9822(8) | 10.314(1) | 6.0916(1) | 108.80(1) | 355.80(1) |
| $Ru_{0.6}Cr_{0.4}Cl_3$ | $Ru_{0.59(1)}Cr_{0.47(1)}Cl_3$ | 5.9696(4) | 10.3380(6) | 6.0869(2) | 108.842(5) | 355.52(2) |
| Ru$_{0.5}$Cr$_{0.5}$Cl$_3$ | Ru$_{0.41(4)}$Cr$_{0.59(4)}$Cl$_3$ | 5.972(1) | 10.328(2) | 6.0955(7) | 108.81(1) | 355.90(1) |
| Ru$_{0.3}$Cr$_{0.7}$Cl$_3$ | Ru$_{0.31(1)}$Cr$_{0.74(1)}$Cl$_3$ | 5.9732(5) | 10.306(2) | 6.1187(3) | 108.76(1) | 356.65(4) |
| CrCl$_3$ | CrCl$_3$ | 5.9588(1) | 10.3206(1) | 6.1138(2) | 108.495(8) | 356.57(1) |

Single crystal X-ray diffraction patterns were collected for the samples with a nominal $x = 0.05$, 0.1, 0.2 and 0.5 and the structure refinements were conducted. Unfortunately, no suitable crystals with $x > 0.5$ could be isolated for structure analysis.

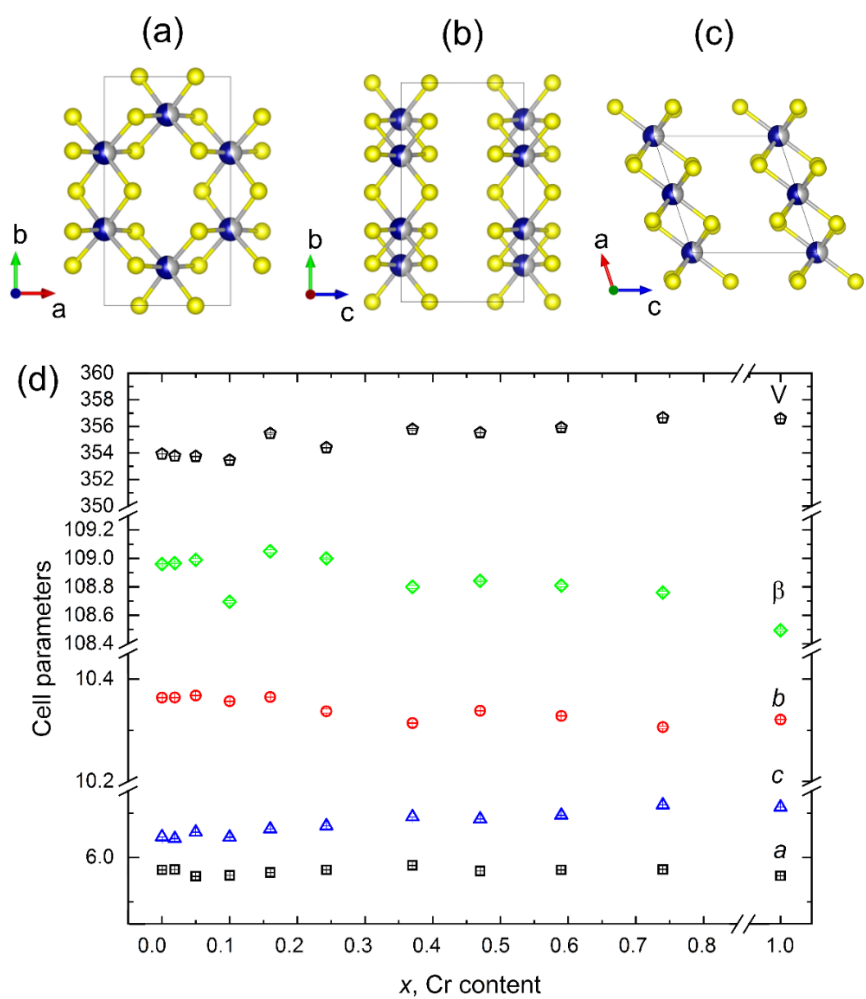

**Figure 3.** Crystal structure and lattice parameters for the Ru$_{1-x}$Cr$_x$Cl$_3$ series. Ru, Cr and Cl atoms are shown in blue, gray and yellow, respectively.

The diffraction images of the investigated crystals of the $Ru_{1-x}Cr_xCl_3$ series resemble those of $α$-$RuCl_3$ and can be indexed with *C*-centered monoclinic unit cells of $a ≈ 5.9$ Å, $b ≈ 10.4$ Å, $c ≈ 6.1$ Å, and $β ≈ 109°$ in agreement with the PXRD data. As observed in [7] reciprocal layers *HK*0, *HK*1, *HK*2, etc. are inconspicuous, but layers perpendicular to the stacking direction of the honeycomb nets show an alteration of rows consisting of sharp Bragg spots only and rows with extended diffuse scattering contributions between Bragg spots or even diffuse intensities instead of Bragg spots. As shown for the example $Ru_{0.91(4)}Cr_{0.10(2)}Cl_3$ in Figure 4, rows 0*KL*, 1*KL*, 2*KL*, etc. contain only sharp Bragg spots for $K = 3n$ (n: integer) and diffuse rods with a modulated intensity distribution for $K = 3n+1$ and $K = 3n+2$. The modulo 3 sequence of the diffuse rods points towards a shift of ±b/3 between stacked honeycomb layers and was called stacking fault of *type a* in [7]. The same "extinction rules" for the location of diffuse scattering rods has been described as a textbook example for stacking faults of hexagonal closed-packed layers. [37] The modulation of the intensity along these rows (Figure 4c) indicates partial ordering, however, this is not resolvable at least with the given setup. Reflections in every third row (00*L*, 03*L*, 06*L*, e.g.) are sharp and without diffuse steaks, i.e., these rows are not affected by stacking faults and the structure factors of the respective Bragg reflections are not altered by atomic shifts. These rows *see* a projected structure which is periodic along the stacking direction with all layers being identical. This, in turn, means it is impossible to probe different Ru/Cr site occupancies (if there were some) in different layers along 001 from X-ray diffraction experiments. On the other hand, the ordering in an individual honeycomb layer can't be resolved either as the scattering experiment averages over several layers. Moreover, the diffuse scattering – which is omitted during conventional integration of reflection intensities – is taking intensities off the Bragg spots, so that the observed structure factors for these rows are subject to large systematic errors. Depending on the amount of diffuse scattering intensity, this will inevitably lead to poor fits, large discrepancies between $F_{obs}$ and $F_{calc}$, large residual electron densities and conspicuous displacement parameters.

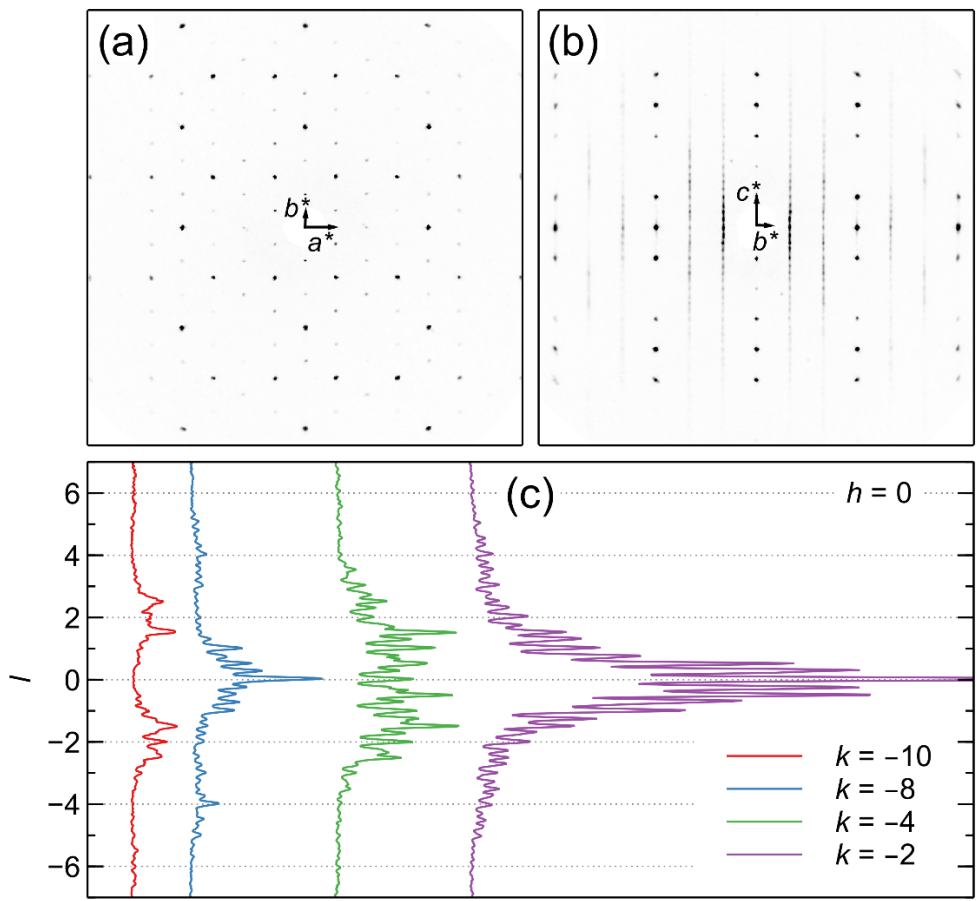

**Figure 4.** Re-calculated reciprocal layers (a) $hk0$ and (b) $0kl$ of $Ru_{0.91(4)}Cr_{0.10(2)}Cl_3$; the intensity distribution along the series containing diffuse contributions exhibit a certain modulation pointing towards partial ordering (c).

Only crystals of the composition $Ru_{0.5}Cr_{0.5}Cl_3$ were found of a quality comparable to pure $\alpha$-$RuCl_3$ [38,7], i.e. with a reasonably low amount of diffuse scattering contributions so that a standard refinement procedure was considered. For samples of $Ru_{1-x}Cr_xCl_3$ no crystals with higher amounts of Cr than $x > 0.5$ and sufficient quality were found.

The structure refinement of $Ru_{0.5}Cr_{0.5}Cl_3$ using the atomic parameters of $\alpha$-$RuCl_3$ as start values and a mixed Ru/Cr occupancy on the metal position (Wyckoff site $4h$) yields a reasonable structure model, however with somehow larger atomic displacement parameters for the metal atoms and a relatively high difference Fourier ($F_{obs}$–$F_{calc}$) maximum of 7 e·Å$^{-1}$ at the $2b$ site (0, ½, 0), in the van der Waals gap directly above and below the voids of the honeycomb layers. For reasons stated above, this maximum does probably not correspond to a new atomic position but is a result of wrong intensities of Bragg reflections caused by diffuse scattering contributions. Omitting all Bragg reflections lying on the diffuse rods $0KL$, $1KL$, $2KL$, with $K = 3n+1$ and $K = 3n+2$, (see next paragraph) the refinement proceeds smoothly to good $R$ values and a featureless difference Fourier map, Table 2. The refinement results in a composition $Ru_{0.48(1)}Cr_{0.52(1)}Cl_3$ close to the nominal one, but slightly richer in the Cr content as derived by EDX ($x = 0.59$). However, the assumed absence of atoms in the van der Waals gaps needs to be cross-checked by local methods, cf. below.

In order to get reasonable refinements and to make at least some estimates about Cr contents and structure features of the $Ru_{1-x}Cr_xCl_3$ samples with $x = 0.05$, 0.1 and 0.2, these data sets were filtered by removing all reflection in rows with diffuse streaks, too. This leads to a drastically decreased number of reflections, but we tried to keep a data/parameter ratio of about 10 to prevent unstable refinements, if necessary by adding restrictions. The results of these refinements are stated in Table 2, Table 3 listed the changes of interatomic distances and angles with $x$. As can clearly be seen, the refined Ru/Cr ratios on the $4h$ site are in line with EDX data and the nominal compositions within analytical limits and the lattice parameters agree with those from powder data. We therefore assume from the X-ray diffraction data, that $Cr^{3+}$ ions substitute $Ru^{3+}$ on its lattice site only. The lattice parameters change slightly upon substitution and the substitution increases the propensity for stacking faults noticeably. The local *M*–Cl coordination geometries reveal only minor change, too, Table 3.

**Table 2.** Crystallographic data refinement results for Ru$_{1-x}$Cr$_x$Cl$_3$ with $x$ = 0.05, 0.1, 0,2 0.5.

| Nominal $x$ | 0.05 | 0.1 | 0.2 | 0.5 | 0.5 |
|---|---|---|---|---|---|
| chemical formula (EDX) | Ru$_{0.96(1)}$Cr$_{0.05(1)}$Cl$_3$ | Ru$_{0.91(4)}$Cr$_{0.10(2)}$Cl$_3$ | Ru$_{0.75(2)}$Cr$_{0(1)}$Cl$_3$ | Ru$_{0.41(4)}$Cr$_{0.59(4)}$Cl$_3$ | |
| refined formula | Ru$_{0.92(1)}$Cr$_{0.52}$Cl$_3$ | Ru$_{0.08(1)}$Cr$_{0.16(1)}$Cl$_3$ | Ru$_{0.75(2)}$Cr$_{0.25(1)}$Cl$_3$ | Ru$_{0.48(1)}$Cr$_{0.52(1)}$Cl$_3$ | Ru$_{0.48(1)}$Cr$_{0.52(1)}$Cl$_3$ |
| formula weight, g/mol | 205.99 | 203.53 | 195.03 | 181.73 | |
| temperature, K | | | 100 | | |
| crystal system, space group, $Z$ | | | monoclinic, *C*2/*m*, 4 | | |
| $a$, Å | 5.9673(5) | 5.9689(8) | 5.9662(8) | 5.9701(15) | |
| $b$, Å | 10.3391(5) | 10.3333(10) | 10.3361(12) | 10.3133(23) | |
| $c$, Å | 6.0301(6) | 6.0422(11) | 6.0300(8) | 6.0496(12) | |
| $\beta$, deg | 109.093(9) | 109.241(18) | 108.756(14) | 108.828(15) | |
| $V$, Å$^3$ | 351.57(5) | 351.86(9) | 352.11(8) | 352.55(14) | |
| $\rho$, g·cm$^{-3}$ | 3.845 | 3.767 | 3.679 | 3.425 | |
| crystal size, mm$^3$ | 0.262×0.102×0.011 | 0.151×0.073×0.017 | 0.112×0.027×0.008 | 0.114×0.082×0.011 | |
| radiation type, wavelength, Å | | | MoK$\alpha$, 0.71073 | | |
| $\mu$, mm$^{-1}$ | 6.36 | 6.26 | 6.14 | 5.81 | |
| rel. transmission | 0.534/1.000 | 0.565/1.000 | 0.635/1.000 | 0.665/1.000 | |
| $\theta_{max}$° | 44.42 | 34.65 | 45.65 | 45.40 | |
| data set | filtered | filtered | filtered | full | filtered |
| data/parameters* | 172/22 | 87/9 | 171/13 | 1529/22 | 172/22 |
| reflections meas./unique/filtered | 13959/1539/1466 | 2050/747/220 | 3690/1523/402 | 8275/1529/– | 8275/1529/918 |
| $R_{int}$ | 0.0402/0.029 | 0.0220/0.018 | 0.0290/0.020 | 0.016 | 0.018 |
| $R_1/wR_2$ * | 0.015/0.036 | 0.031/0.081 | 0.028/0.062 | 0.0236/0.0505 | 0.013/0.027 |
| *GooF*\* | 1.105 | 1.233 | 1.122 | 1.110 | 1.262 |
| $\Delta\rho_{max}/\Delta\rho_{min}$, e Å$^{-1}$ | 0.33/–0.64 | 0.88/–0.66 | 0.52/–0.99 | 7.81/–0.89 | 0.314/–0.267 |

*: for all reflections in the actual refinement

**Table 3.** Selected structure data for Ru$_{1-x}$Cr$_x$Cl$_3$ with $x$ = 0, 0.05, 0.1, 0,2 0.5 (filtered data) as well as for CrCl$_3$.[39]

| Nominal $x$ | 0 | 0.05 | 0.1 | 0.2 | 0.5 | 1 |
|---|---|---|---|---|---|---|
| d ($M$–Cl1) / Å | 2.357(1) | 2.380(1) | 2.352(1) | 2.335(1) | 2.361(1) | 2.347(1) |
| d ($M$–Cl2) / Å | 2.354(1) / 2.356(1) | 2.379(1) /2.420(1) | 2.367(1) / 2.371(1) | 2.350(1) / 2.374(1) | 2.346(1) / 2.352(1) | 2.342(1) / 2.340(1) |
| Angle (Cl1–$M$–Cl2) / ° | 91.11(1) | 88.22(1) | 90.80(1) | 92.41(1) | 92.13(1) | 93.10(1) |

**TEM investigations**

Since the XRD analysis provides only limited information about the stacking sequence in the crystals, a TEM study has been undertaken using the Ru$_{0.91(4)}$Cr$_{0.10(2)}$Cl$_3$ sample. EDX analysis performed on lamellae as well as on several crystallites confirms the composition of the TEM specimen.

Figure 5 provides typical SAED patterns collected by tilting the sample to the angles corresponding to the desired zone axes, following Kikuchi lines. The reflection conditions $hkl$: $h+k=2n$; $h0l$: $h=2n$; $0kl$: $k=2n$; $hk0$: $h+k=2n$; $0k0$: $k=2n$; $h00$: $h=2n$ confirm the $C2/m$ space group. The lattice parameters derived from the ED patterns are in line with the XRD data $a \approx 5.9$ Å, $b \approx 10.4$ Å, $c \approx 6.1$ Å, $\beta \approx 109°$. In both 001 and 010 zone-axis, no particular disorder is visible. However, the structure exhibits a diffuse streak along the $c$-direction in its 110 and 100 zone-axis indicating a strong interlayer stacking disorder revealed also by XRD.

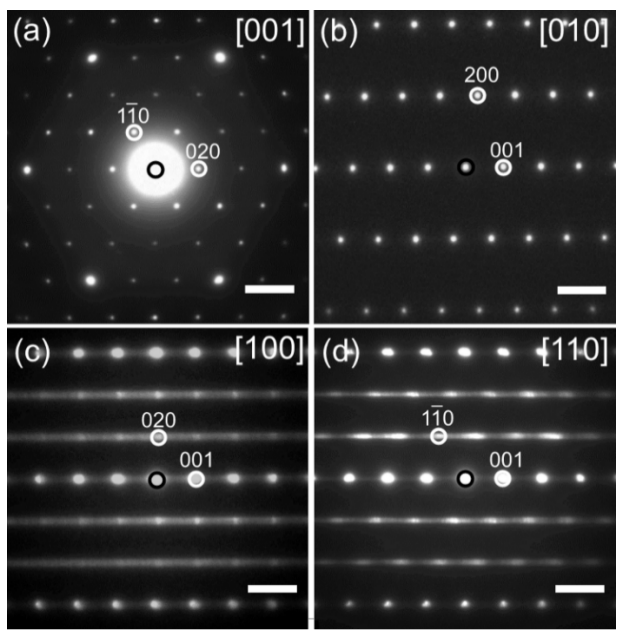

**Figure 5.** Electron diffraction patterns of a $Ru_{0.91(4)}Cr_{0.10(2)}Cl_3$ sample along the (a) 001, (b) 010, (c) 100, and (d) 110 zone-axis directions. The scale bar corresponds to 2 $nm^{-1}$.

Figure 6 presents a TEM image of the sample recorded along the 001 zone axis. In this projection, a distorted hexagonal pattern is formed by the voids between the Ru/Cr atomic columns. The distance between the centers of these voids is about 5.9 Å. Thus, the mixed-metal compound preserves the honeycomb type lattice of α-$RuCl_3$ in the *ab*-plane, with no voids filling.

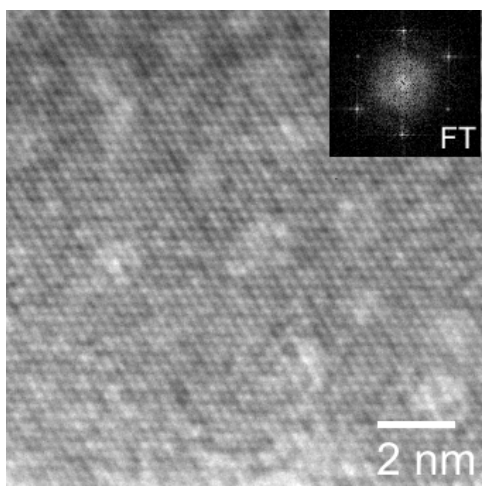

**Figure 6.** TEM image of $Ru_{0.91(4)}Cr_{0.10(2)}Cl_3$ sample taken in the 001 zone-axis and the corresponding FFT.

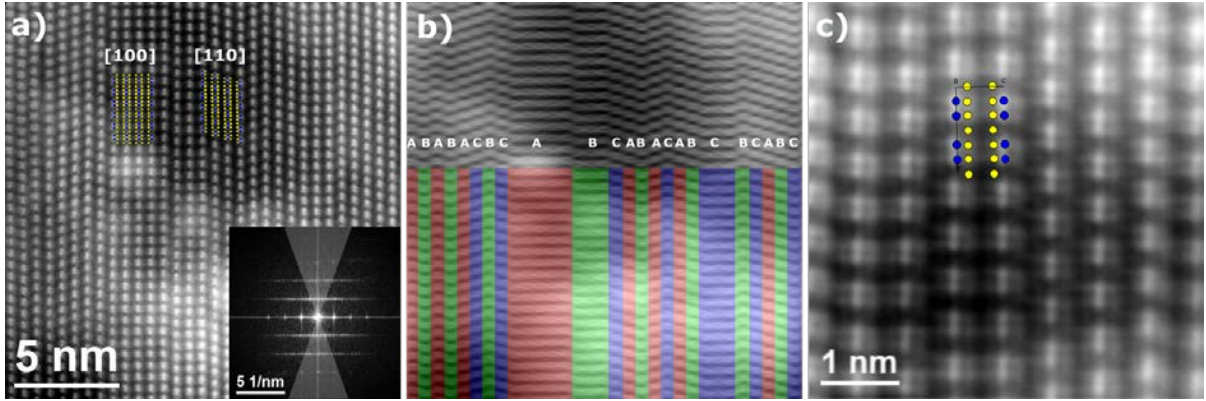

**Figure 7.** (a) HR-STEM image of the $Ru_{0.91(4)}Cr_{0.10(2)}Cl_3$ sample taken along the 100 zone-axis with overlaid atomic structure for the 100 and 110 zone axis. Inset shows the FFT of the image and the marked mask for filtered image b. (b) Fourier filtered image of a, with color coded and marked stacking sequences. (c) Close-up of the atomic structure with of an atomic overlay of the 100 zone axis, blue and yellow circles denote Ru/Cr and Cl columns, respectively.

To further elaborate the origin of the observed streaks in the ED pattern in 100 zone axis orientation, HR-STEM images have been acquired. Figure 7 shows the atomic resolution images, where due to the high atomic number, the Ru/Cr atom columns are clearly visible. Even though the sample was oriented in the 100 zone axis, areas which resemble the 110 atomic arrangement are found. To further analyze this behavior, Fourier filtered images using the wedge mask shown in Figure 7a-inset are created (see Figure 7b). Here, the local stacking (A, B, or C) of the whole image was analyzed and color coded. Stacking sequences of e.g. AAA… are assigned to 100 oriented areas, whereas stacking sequences of ABC… are assigned to 110 oriented areas. No clear long range stacking sequence could be found, resulting in the observed streaks in the ED pattern in 100 and 110 zone axis orientation. Analyzing the atomic stacking in detail (see Figure 7c), we found that only shifts between the layers corresponding to the Cl⋯Cl van der Waals gap are observable.

It is worth mentioning that stacking faults in pure α-RuCl$_3$ may affect the Heisenberg and asymmetric exchange interactions and may, thus, be a reason of a long-range magnetic ordering at low temperatures[40,41]. For α-RuCl$_3$, e.g., specific heat and susceptibility measurements reveal two separate Néel transitions at about 7 and 14 K on moderate-quality single crystals [42,43,29], while highest-quality crystals exhibit only the phase transition at the lower temperature. [38,44] In the case of the Ru$_{1-x}$Cr$_x$Cl$_3$ sample with $x = 0.1$ the magnetic susceptibility measurements revealed that instead of two Néel transitions only a broad hump is observed near the antiferromagnetic ordering temperature (see [45] and Figure 9 in Magnetic properties measurements part below). We assume that the broadening of the magnetic transitions may reflect the increased disorder along $c$-axis as compared to pure ABAB-stacked α-RuCl$_3$, as revealed by HR-STEM. Due to sensitivity to the local electronic environment in 2D systems, disorder might have a profound impact that masks the intrinsic magnetic behavior. Thus, the stacking sequence of the layers, as well as small distortions inside each layer are important for the understanding and a consistent description of the physical properties of Ru$_{1-x}$Cr$_x$Cl$_3$.

**Raman spectroscopy**

Figure 8 shows the non-polarized Raman spectra of Ru$_{1-x}$Cr$_x$Cl$_3$ samples with $x = 0$, 0.02, 0.05, 0.1, 0.5, and 1 measured using the 633 nm excitation line at $T = 300$ K on single crystals. The spectrum of pure α-RuCl$_3$ and CrCl$_3$ reproduce the literature data for the anhydrous metal trichlorides [46,47,48,49] and at least five phonon modes can be observed in accordance with factor group analysis for the $C2/m$ space group. These modes can be identified as $A_g + B_g$ doublets which cannot be resolved with the given experimental setup. The Raman spectra of the ternary samples can be fully described as a superposition of the corresponding RuCl$_3$ and CrCl$_3$ spectra. With increasing chromium content up to $x = 0.02$, the CrCl$_3$ mode at 247.9 cm$^{-1}$ becomes visible. For the sample with $x = 0.1$, modes at 117.4, 166.3, 209 and 345.5 cm$^{-1}$ are evident. Thus, the apparent lack of additional Raman-active lines is assigned to the symmetry preservation in the composition range Ru$_{1-x}$Cr$_x$Cl$_3$ ($0.02 \leq x \leq 0.5$) which is in line with our XRD and TEM data. The incorporation of chromium into α-RuCl$_3$ only modifies bond lengths and reduced masses and, hence, the exchange parameters within the same lattice type. Thus, the effects of the investigated substitution are restricted to modifications of bond lengths and reduced masses and, hence, the exchange parameters within the same lattice type. The Raman data bear no evidence for a change of the oxidation state of Ru$^{3+}$ as it has been observed upon sodium intercalation. [50]

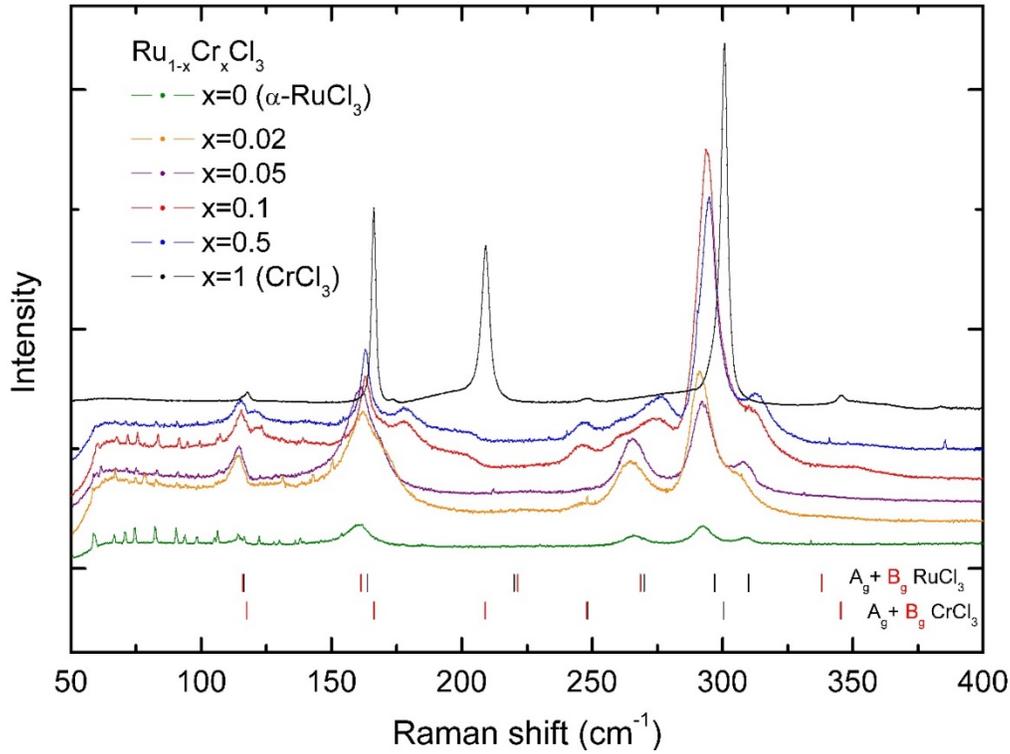

**Figure 8.** Raman spectra of the Ru$_{1-x}$Cr$_x$Cl$_3$ series recorded with an excitation wavelength of 633 nm.

**Magnetic properties measurements**

The temperature dependence of the magnetic susceptibility of Ru$_{1-x}$Cr$_x$Cl$_3$ single crystals in applied magnetic fields $H \parallel ab$ is represented in Figure 9. For the samples with $x = 0.1$ and $0.5$ temperature spin-glass state is observed. Furthermore, upon the series the room temperature magnetic susceptibility increases progressively with increasing Cr content $x$, which is consistent with the addition of $s = 3/2$ Cr$^{3+}$ moments into the $J_{\text{eff}} = \frac{1}{2}$ Ru$^{3+}$ matrix. The low temperature magnetic susceptibility increases over two orders of magnitude from $\alpha$-RuCl$_3$ to CrCl$_3$, and which is mainly attributed to ferromagnetic interactions between neighboring Cr sites within the honeycomb layers.

A Curie-Weiss fit in the paramagnetic high-temperature region between 200 K and 300 K yields large effective paramagnetic moments $\mu_{eff}$ = 2.2(2) $\mu_B$ of the $J_{eff}$ = 1/2 ruthenium ions in α-RuCl$_3$ due to its mixed spin-orbital character [5], while the effective moment increases towards $\mu_{eff}$ = 3.5(3) $\mu_B$, which is slightly lower than the expected spin-only value $\mu_{eff}$ = 3.87 $\mu_B$ for the $s$ = 3/2 chromium ions in CrCl$_3$.

In the temperature region below about 10 K, the magnetic susceptibility of α-RuCl$_3$ shows an antiferromagnetic transition into a zigzag order at $T_N$ = 7 K and a shoulder around $T_N$ = 10 K. According to previous studies, this indicates a dominant ABC stacking in the sample with a small amount of stacking faults [12]. A splitting between the zero-field cooled and the field-cooled magnetization at low temperatures for $x$ = 0.1 and $x$ = 0.5 suggests that the antiferromagnetic ground state of α-RuCl$_3$ could be changed upon Cr doping either into a canted antiferromagnetic structure or into a spin-glass state. The magnetic susceptibility of CrCl$_3$ then indicates the 3D antiferromagnetic transition at $T_N$ = 14 K with a preceding strong increase at slightly higher temperatures indicating the ferromagnetic 2D order within the honeycomb planes at $T_C$ ≈ 18 K in good agreement with previous studies [20, 51]. In general, our magnetic measurements confirm there are no impurity phases or phase separations in the Ru$_{1-x}$Cr$_x$Cl$_3$ crystals under investigation.

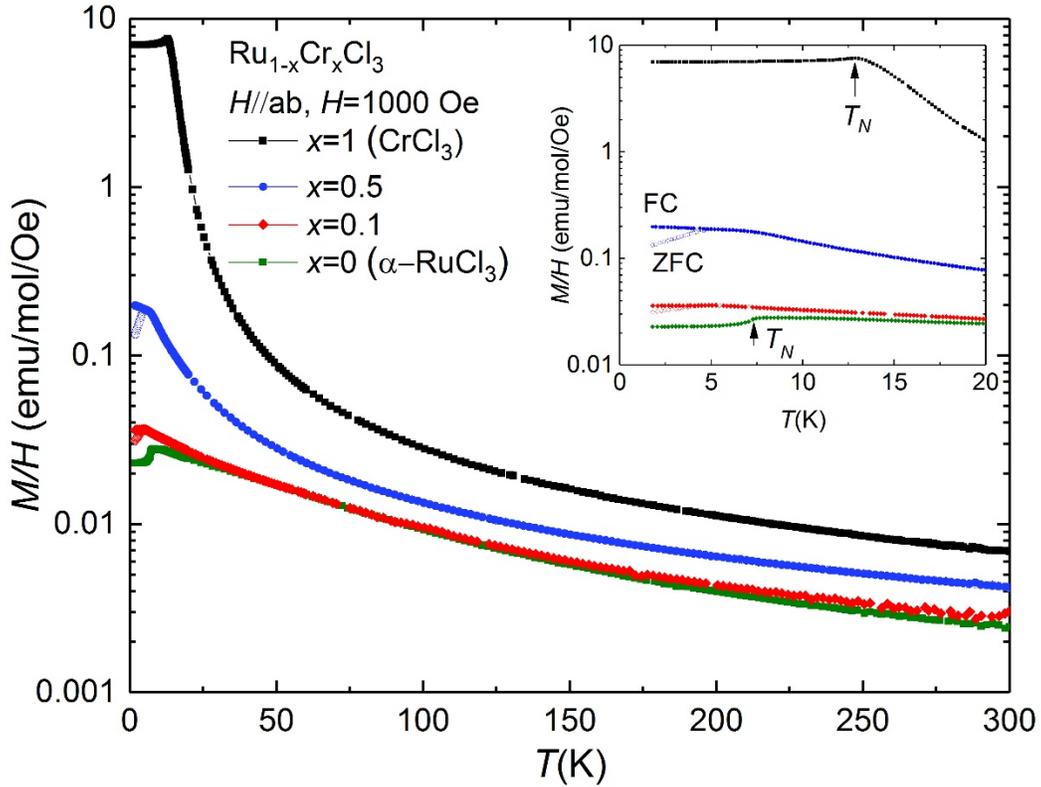

**Figure 9.** Temperature dependence of the normalized magnetization *M/H* of $Ru_{1-x}Cr_xCl_3$ on a semi-logarithmic scale. The magnetic field of 1000 Oe was applied in the *ab* plane of $Ru_{1-x}Cr_xCl_3$ single crystals. The inset is a zoom into the low-temperature region. Open and full symbols stand for zero-field-cooled (ZFC) and field-cooled (FC) magnetization measurements.

**Summary**

*α*-RuCl₃ serves as a host matrix for magnetic $Cr^{3+}$ dopant. Single crystals of the materials $Ru_{1-x}Cr_xCl_3$ ($0 \leq x \leq 1$) were grown by chemical vapor transport reaction. Crystallographic studies reveal that the mixed-metal compounds preserve the honeycomb type lattice of α-RuCl₃ with mixed occupancy of Ru/Cr sites. The versatility of the doping allows for extensive magnetic, microscopic and spectroscopic measurements in a broad range of chromium concentrations. The ground state of α-RuCl₃ can be changed upon Cr doping either into a canted antiferromagnetic structure or into a spin-glass state by emerging Cr–Ru exchange interactions. These results contribute to the fundamental experimental understanding of the effect of magnetic impurities in Kitaev materials beyond $\alpha$-RuCl$_3$ and exhibit one of possible chemistry routes for the fabrication of devices for spin-orbitronic and spin-photonic applications.

ASSOCIATED CONTENT

**Supporting Information**. The following files are available free of charge.

CIF files are available from the Cambridge Structural Database (CSD):

Ru$_{0.5}$Cr$_{0.5}$Cl$_3$: CSD-1885667 (full data set, unfiltered)

Ru$_{0.95}$Cr$_{0.05}$Cl$_3$: CSD-1885668 (filtered data set)

Ru$_{0.9}$Cr$_{0.1}$Cl$_3$: CSD-1885669 (filtered)

Ru$_{0.8}$Cr$_{0.2}$Cl$_3$: CSD-1885687 (filtered).

XRD patterns for all Ru$_{1-x}$Cr$_x$Cl$_3$ ($0 \leq x \leq 1$) samples, a SEM image of the lamellae micromachined from the Ru$_{0.91(4)}$Cr$_{0.10(2)}$Cl$_3$ crystal, typical BF and DF low-resolution images of regions with and without Ga implantation and corresponding EDX spectra (PDF).

AUTHOR INFORMATION

**Corresponding Author**

*maria.roslova@mmk.su.se

**Present Addresses**

† Maria Roslova is currently affiliated to the Stockholm University, Svante Arrhenius väg 16C, SE-106 91 Stockholm, Sweden.

**Author Contributions**

The manuscript was written through contributions of all authors. All authors have given approval to the final version of the manuscript.

**Funding Sources**

DFG, SFB-1143.

ACKNOWLEDGMENT

We cordially thank A. Brünner and M. Münch for technical support, Ch. Damm for the introduction to Tecnai G2 microscope, A. Pohl and D. Bieberstein for the FIB lamella preparation. We thank Prof. J. J. Weigand for providing measuring time at the SuperNova single crystal X-ray diffractometer (Rigaku-Oxford Diffraction). Financial support by the Deutsche Forschungsgemeinschaft in the framework of the SFB-1143 "Correlated Magnetism – From Frustration to Topology" (projects B01 and B03) is acknowledged.

REFERENCES

(1)     Kitaev, A. Anyons in an Exactly Solved Model and Beyond. *Ann. Phys.* **2006**, *321* (1), 2–111. https://doi.org/10.1016/j.aop.2005.10.005.

(2)     Nayak, C.; Simon, S. H.; Stern, A.; Freedman, M.; Das Sarma, S. Non-Abelian Anyons and Topological Quantum Computation. *Rev. Mod. Phys.* **2008**, *80* (3), 1083–1159. https://doi.org/10.1103/RevModPhys.80.1083.

(3) Kitaev, A.; Laumann, C. Topological Phases and Quantum Computation. *arXiv:0904.2771 [cond-mat.mes-hall]*. **2009**.

(4) Pachos, J. K. *Introduction to Topological Quantum Computation*; Cambridge University Press, 2012.

(5) Do, S.-H.; Park, S.-Y.; Yoshitake, J.; Nasu, J.; Motome, Y.; Kwon, Y. S.; Adroja, D. T.; Voneshen, D. J.; Kim, K.; Jang, T.-H.; et al. Incarnation of Majorana Fermions in Kitaev Quantum Spin Lattice. *arXiv:1703.01081 [cond-mat.str-el]*. **2017**.

(6) Narozhny, B. Majorana Fermions in the Nonuniform Ising-Kitaev Chain: Exact Solution. *Sci. Rep.* **2017**, *7* (1), 1447. https://doi.org/10.1038/s41598-017-01413-z.

(7) Johnson, R. D.; Williams, S. C.; Haghighirad, A. A.; Singleton, J.; Zapf, V.; Manuel, P.; Mazin, I. I.; Li, Y.; Jeschke, H. O.; Valentí, R.; et al. Monoclinic Crystal Structure of α-RuCl$_3$ and the Zigzag Antiferromagnetic Ground State. *Phys. Rev. B* **2015**, *92* (23), 235119. https://doi.org/10.1103/PhysRevB.92.235119.

(8) M. Fletcher, J.; E. Gardner, W.; C. Fox, A.; Topping, G. X-Ray, Infrared, and Magnetic Studies of α- and β-Ruthenium Trichloride. *J. Chem. Soc. Inorg. Phys. Theor.* **1967**, *0* (0), 1038–1045. https://doi.org/10.1039/J19670001038.

(9) Fletcher, J. M.; Gardner, W. E.; Hooper, E. W.; Hyde, K. R.; Moore, F. H.; Woodhead, J. L. Anhydrous Ruthenium Chlorides. *Nature* **1963**, *199* (4898), 1089–1090. https://doi.org/10.1038/1991089a0.

(10) Stroganov E. V.; Ovchinnikov K. V. Crystal Structure of Ruthenium Trichloride. *Vestn. Leningr. Univ Ser Fiz Khim* **1957**, *12*, 152.

(11) Brodersen, K.; Thiele, G.; Ohnsorge, H.; Recke, I.; Moers, F. Die Struktur Des IrBr$_3$ Und Über Die Ursachen Der Fehlordnungserscheinungen Bei Den in Schichtenstrukturen

Kristallisierenden Edelmetalltrihalogeniden. *J. Common Met.* **1968**, *15* (3), 347–354. https://doi.org/10.1016/0022-5088(68)90194-X.

(12) Cao, H. B.; Banerjee, A.; Yan, J.-Q.; Bridges, C. A.; Lumsden, M. D.; Mandrus, D. G.; Tennant, D. A.; Chakoumakos, B. C.; Nagler, S. E. Low-Temperature Crystal and Magnetic Structure of α-RuCl$_3$. *Phys. Rev. B* **2016**, *93* (13), 134423. https://doi.org/10.1103/PhysRevB.93.134423.

(13) Ketelaar, J. a. A.; MacGillavry, C. H.; Renes, P. A. The Crystal Structure of Aluminium Chloride. *Recl. Trav. Chim. Pays-Bas* **1947**, *66* (8), 501–512. https://doi.org/10.1002/recl.19470660805.

(14) Choi, S. K.; Coldea, R.; Kolmogorov, A. N.; Lancaster, T.; Mazin, I. I.; Blundell, S. J.; Radaelli, P. G.; Singh, Y.; Gegenwart, P.; Choi, K. R.; et al. Spin Waves and Revised Crystal Structure of Honeycomb Iridate Na$_2$IrO$_3$. *Phys. Rev. Lett.* **2012**, *108* (12), 127204. https://doi.org/10.1103/PhysRevLett.108.127204.

(15) O'Malley, M. J.; Verweij, H.; Woodward, P. M. Structure and Properties of Ordered Li$_2$IrO$_3$ and Li$_2$PtO$_3$. *J. Solid State Chem.* **2008**, *181* (8), 1803–1809. https://doi.org/10.1016/j.jssc.2008.04.005.

(16) Lampen-Kelley, P.; Banerjee, A.; Aczel, A. A.; Cao, H. B.; Stone, M. B.; Bridges, C. A.; Yan, J.-Q.; Nagler, S. E.; Mandrus, D. Destabilization of Magnetic Order in a Dilute Kitaev Spin Liquid Candidate. *Phys. Rev. Lett.* **2017**, *119* (23), 237203. https://doi.org/10.1103/PhysRevLett.119.237203.

(17) Ludwig, T. Struktur-Eigenschafts-Beziehungen von Übergangsmetalltrihalogeniden mit niederdimensionalen Bindungssystemen, Albert-Ludwigs-Universität: Freiburg, 2001.

(18) Schmidt, P. J. Untersuchung Der Elektronischen Struktur von Mischkristallen Mit Übergangsmetallhalogeniden, Albert-Ludwigs-Universität: Freiburg, 1999.

(19) Hillebrecht, H.; Schmidt, P. J.; Rotter, H. W.; Thiele, G.; Zönnchen, P.; Bengel, H.; Cantow, H.-J.; Magonov, S. N.; Whangbo, M.-H. Structural and Scanning Microscopy Studies of Layered Compounds $MCl_3$ (M = Mo, Ru, Cr) and $MOCl_2$ (M = V, Nb, Mo, Ru, Os). *J. Alloys Compd.* **1997**, *246* (1), 70–79. https://doi.org/10.1016/S0925-8388(96)02465-6.

(20) McGuire, M. A.; Clark, G.; KC, S.; Chance, W. M.; Jellison, G. E.; Cooper, V. R.; Xu, X.; Sales, B. C. Magnetic Behavior and Spin-Lattice Coupling in Cleavable van Der Waals Layered $CrCl_3$ Crystals. *Phys. Rev. Mater.* **2017**, *1* (1), 014001. https://doi.org/10.1103/PhysRevMaterials.1.014001.

(21) Cable, J. W.; Wilkinson, M. K.; Wollan, E. O. Neutron Diffraction Investigation of Antiferromagnetism in $CrCl_3$. *J. Phys. Chem. Solids* **1961**, *19* (1), 29–34. https://doi.org/10.1016/0022-3697(61)90053-1.

(22) Binnewies, M.; Glaum, R.; Schmidt, M.; Schmidt, P. *Chemical Vapor Transport Reactions*; Walter de Gruyter, 2012.

(23) Oppermann, H. Das Reaktionsgleichgewicht 2 $CrCl_{3f,g}$ + $Cl_{2g}$ = 2 $CrCl_{4g}$. Mit 5 Abbildungen. *Z. Für Anorg. Allg. Chem.* **1968**, *359* (1–2), 51–57. https://doi.org/10.1002/zaac.19683590107.

(24) Plies, V. Massenspektrometrische Untersuchungen Der Gasphase Über $CrCl_3$ Und $CrCl_3/Cl_2$. *Z. Für Anorg. Allg. Chem.* **1988**, *556* (1), 120–128. https://doi.org/10.1002/zaac.19885560112.

(25) *APEX2, Version 2014/9*; Bruker AXS Inc., Madison, Wisconsin, USA, 2014.

(26) *CrysAlisPRO*; CrysAlisPRO, Oxford Diffraction /Agilent Technologies UK Ltd, Yarnton, UK, 2015.

(27) Petříček, V.; Dušek, M.; Palatinus, L. Crystallographic Computing System JANA2006: General Features. *Z. Für Krist. - Cryst. Mater.* **2014**, *229* (5), 345–352. https://doi.org/10.1515/zkri-2014-1737.

(28) Jones, L.; Yang, H.; Pennycook, T. J.; Marshall, M. S. J.; Van Aert, S.; Browning, N. D.; Castell, M. R.; Nellist, P. D. Smart Align—a New Tool for Robust Non-Rigid Registration of Scanning Microscope Data. *Adv. Struct. Chem. Imaging* **2015**, *1* (1), 8. https://doi.org/10.1186/s40679-015-0008-4.

(29) Sears, J. A.; Songvilay, M.; Plumb, K. W.; Clancy, J. P.; Qiu, Y.; Zhao, Y.; Parshall, D.; Kim, Y.-J. Magnetic Order in $\alpha$-RuCl$_3$: A Honeycomb-Lattice Quantum Magnet with Strong Spin-Orbit Coupling. *Phys. Rev. B* **2015**, *91* (14), 144420. https://doi.org/10.1103/PhysRevB.91.144420.

(30) He, M.; Wang, X.; Wang, L.; Hardy, F.; Wolf, T.; Adelmann, P.; Brückel, T.; Yixi Su; Meingast, C. Uniaxial and Hydrostatic Pressure Effects in $\alpha$-RuCl$_3$ Single Crystals via Thermal-Expansion Measurements. *J. Phys. Condens. Matter* **2018**, *30* (38), 385702. https://doi.org/10.1088/1361-648X/aada1e.

(31) Yang, S.; Kang, J.; Yue, Q.; Yao, K. Vapor Phase Growth and Imaging Stacking Order of Bilayer Molybdenum Disulfide. *J. Phys. Chem. C* **2014**, *118* (17), 9203–9208. https://doi.org/10.1021/jp500050r.

(32) Atuchin, V. V.; Borisov, S. V.; Gavrilova, T. A.; Kokh, K. A.; Kuratieva, N. V.; Pervukhina, N. V. Physical Vapor Transport Growth and Morphology of Bi$_2$Se$_3$

Microcrystals. *Particuology* **2016**, *26*, 118–122. https://doi.org/10.1016/j.partic.2015.10.003.

(33) Ma, X.-H.; Cho, K.-H.; Sung, Y.-M. Growth Mechanism of Vertically Aligned SnSe Nanosheets via Physical Vapour Deposition. *CrystEngComm* **2014**, *16* (23), 5080–5086. https://doi.org/10.1039/C4CE00213J.

(34) Grönke, M.; Schmidt, P.; Valldor, M.; Oswald, S.; Wolf, D.; Lubk, A.; Büchner, B.; Hampel, S. Chemical Vapor Growth and Delamination of α-RuCl$_3$ Nanosheets down to the Monolayer Limit. *Nanoscale* **2018**, *10* (40), 19014–19022. https://doi.org/10.1039/C8NR04667K.

(35) Lampen-Kelley, P.; Banerjee, A.; Aczel, A. A.; Cao, H. B.; Yan, J.-Q.; Nagler, S. E.; Mandrus, D. Destabilization of Magnetic Order in a Dilute Kitaev Spin Liquid Candidate. *arXiv:1612.07202 [cond-mat.str-el]*. **2016**.

(36) Shannon, R. D. Revised Effective Ionic Radii and Systematic Studies of Interatomic Distances in Halides and Chalcogenides. *Acta Crystallogr. A* **1976**, *32* (5), 751–767. https://doi.org/10.1107/S0567739476001551.

(37) Neder, R. B.; Proffen, T. *Diffuse Scattering and Defect Structure Simulations: A Cook Book Using the Program DISCUS*; Oxford University Press, 2008.

(38) Hentrich, R.; Wolter, A. U. B.; Zotos, X.; Brenig, W.; Nowak, D.; Isaeva, A.; Doert, T.; Banerjee, A.; Lampen-Kelley, P.; Mandrus, D. G.; et al. Large Field-Induced Gap of Kitaev-Heisenberg Paramagnons in α -RuCl$_3$ *Phys. Rev. Lett.* **2018**, *120, 117204*. https://doi.org/10.1103/PhysRevLett.120.117204.

(39) Morosin, B.; Narath, A. X-Ray Diffraction and Nuclear Quadrupole Resonance Studies of Chromium Trichloride. *J. Chem. Phys.* **1964**, *40* (7), 1958–1967. https://doi.org/10.1063/1.1725428.

(40) Yadav, R.; Bogdanov, N. A.; Katukuri, V. M.; Nishimoto, S.; Brink, J. van den; Hozoi, L. Kitaev Exchange and Field-Induced Quantum Spin-Liquid States in Honeycomb α-RuCl$_3$. *Sci. Rep.* **2016**, *6*, srep37925. https://doi.org/10.1038/srep37925.

(41) Winter, S. M.; Li, Y.; Jeschke, H. O.; Valentí, R. Challenges in Design of Kitaev Materials: Magnetic Interactions from Competing Energy Scales. *Phys. Rev. B* **2016**, *93* (21), 214431. https://doi.org/10.1103/PhysRevB.93.214431.

(42) Majumder, M.; Schmidt, M.; Rosner, H.; Tsirlin, A. A.; Yasuoka, H.; Baenitz, M. Anisotropic $Ru^{3+}$ $4d^5$ Magnetism in the α-RuCl$_3$ Honeycomb System: Susceptibility, Specific Heat, and Zero-Field NMR. *Phys. Rev. B* **2015**, *91* (18), 180401. https://doi.org/10.1103/PhysRevB.91.180401.

(43) Banerjee, A.; Bridges, C. A.; Yan, J.-Q.; Aczel, A. A.; Li, L.; Stone, M. B.; Granroth, G. E.; Lumsden, M. D.; Yiu, Y.; Knolle, J.; et al. Proximate Kitaev Quantum Spin Liquid Behaviour in a Honeycomb Magnet. *Nat. Mater.* **2016**, *Advance online publication*. https://doi.org/10.1038/nmat4604.

(44) Wolter, A. U. B.; Corredor, L. T.; Janssen, L.; Nenkov, K.; Schönecker, S.; Do, S.-H.; Choi, K.-Y.; Albrecht, R.; Hunger, J.; Doert, T.; et al. Field-Induced Quantum Criticality in the Kitaev System α-RuCl$_3$. *Phys. Rev. B* **2017**, *96* (4), 041405. https://doi.org/10.1103/PhysRevB.96.041405.

(45) Bastien, G.; Roslova, M.; Haghighi, M. H.; Mehlawat, K.; Hunger, J.; Isaeva, A.; Doert, T.; Wolter, A. U. B.; Buchner, B. Spin Glass Ground State and Reversed Magnetic

Anisotropy Induced by Cr Doping in the Kitaev Heisenberg Magnet α-RuCl$_3$. *In preparation*. **2018**.

(46) Glamazda, A.; Lemmens, P.; Do, S.-H.; Kwon, Y. S.; Choi, K.-Y. Relation between Kitaev Magnetism and Structure in α-RuCl$_3$. *Phys. Rev. B* **2017**, *95* (17), 174429. https://doi.org/10.1103/PhysRevB.95.174429.

(47) Bermudez, V. M. Unit-Cell Vibrational Spectra of Chromium Trichoride and Chromium Tribromide. *Solid State Commun.* **1976**, *19* (8), 693–697. https://doi.org/10.1016/0038-1098(76)90899-1.

(48) Borghesi, A.; Guizzetti, G.; Marabelli, F.; Nosenzo, L.; Reguzzoni, E. Far-Infrared Optical Properties of CrCl$_3$ and CrBr$_3$. *Solid State Commun.* **1984**, *52* (4), 463–465. https://doi.org/10.1016/0038-1098(84)90036-X.

(49) Avram, C. N.; Gruia, A. S.; Brik, M. G.; Barb, A. M. Calculations of the Electronic Levels, Spin-Hamiltonian Parameters and Vibrational Spectra for the CrCl$_3$ Layered Crystals. *Phys. B Condens. Matter* **2015**, *478*, 31–35. https://doi.org/10.1016/j.physb.2015.08.025.

(50) Koitzsch, A.; Habenicht, C.; Müller, E.; Knupfer, M.; Büchner, B.; Kandpal, H. C.; van den Brink, J.; Nowak, D.; Isaeva, A.; Doert, T. J$_{eff}$ Description of the Honeycomb Mott Insulator α−RuCl$_3$. *Phys. Rev. Lett.* **2016**, *117* (12), 126403. https://doi.org/10.1103/PhysRevLett.117.126403.

(51) Kuhlow, B. Magnetic Ordering in CrCl$_3$ at the Phase Transition. *Phys. Status Solidi A* **1982**, *72* (1), 161–168. https://doi.org/10.1002/pssa.2210720116.